\documentclass{new_tlp}
\usepackage{amssymb}
\usepackage{amsmath}
\usepackage{url}
\usepackage{color}
\usepackage{textcomp}

\usepackage[ruled,vlined]{algorithm2e}
\usepackage{listings}
\usepackage{fancyvrb}
\usepackage{subcaption}
\usepackage{IEEEtrantools}
\usepackage{enumitem}
\usepackage{krudces-main}
\usepackage{krudces-thm}
\usepackage{krudces-thme}

\usepackage{krudces-notation}

\definecolor{darkred}{rgb}{0.8,0,0.2}

\def\K{\mathbf{K}\, }
\def\M{\mathbf{M}\, }
\let\sneg\relax
\newcommand{\sneg}{\ensuremath{\text{-}}}
\def\eclingo{{\tt eclingo}}
\def\wviews{{\tt Wviews}}

\def\AS{\text{\rm AS}}
\newcommand\wv{\mathbb{W}}

\def\Head{\mathit{Head}}

\def\bL{\K}
\def\clingo{{\tt clingo}}
\def\gringo{{\tt gringo}}
\def\wviews{{\tt Wviews}}
\def\EPASP{{\tt EP-ASP}}
\def\eligible{\mathit{eligible}}
\def\minority{\mathit{minority}}
\def\high{\mathit{high}}
\def\fair{\mathit{fair}}
\def\interview{\mathit{interview}}

\def\Facts{\mathit{Facts}}
\def\Heads{\mathit{Heads}}

\begin{document}

\submitted{}
\revised{}
\accepted{}
\title[A solver for Epistemic Logic Programs]
    {\eclingo: A solver for Epistemic Logic Programs\thanks{Partially supported by MINECO, Spain, grant TIC2017-84453-P.
    The second author is funded by the Alexander von Humboldt Foundation.}}

  \author[P. Cabalar, J. Fandinno, J. Garea, J. Romero and T. Schaub]
         {Pedro Cabalar$^1$, Jorge Fandinno$^2$, Javier Garea$^1$, Javier Romero$^2$ and Torsten Schaub$^2$\\
          $^1$ University of Corunna, Spain.\\
          {\email{\{cabalar,javier.garea\}@udc.es}}\\\\
          $^2$ University of Potsdam, Germany\\
          {\email{\{fandinno,javier,torsten\}@uni-potsdam.de}}\\
        }

%\volume{\textbf{10} (3):}
%\jdate{month*** 2006}
%\setcounter{page}{1}
%\pubyear{2006}

%\input{letter.tex}
\maketitle

\begin{abstract}
We describe \eclingo, a solver for epistemic logic programs under Gelfond 1991 semantics built upon the Answer Set Programming system \clingo.
The input language of \eclingo\ uses the syntax extension capabilities of \clingo{} to define subjective literals that, as usual in
epistemic logic programs, allow for checking the truth of a regular literal in all or in some of the answer sets of a program.
The \eclingo\ solving process follows a guess and check strategy.
It first generates potential truth values for subjective literals and, in a second step, it checks the obtained result with respect to the cautious and brave consequences of the program.
This process is implemented using the multi-shot functionalities of \clingo.
We have also implemented some optimisations, aiming at reducing the search space and, therefore, increasing \eclingo's efficiency in some scenarios.
Finally, we compare the efficiency of \eclingo\ with two state-of-the-art solvers for epistemic logic programs on a pair of benchmark scenarios
and show that \eclingo\ generally outperforms their obtained results.
\end{abstract}

%%%%%%%%%%%%%%%%%%%%%%%%%%%%%%%%%%%%%%%%%%%%%%%%%%%%%%%%%%%%%%%%%%%%%%%%%%%%%%%
\section{Introduction}
\label{sec:intro}
%%%%%%%%%%%%%%%%%%%%%%%%%%%%%%%%%%%%%%%%%%%%%%%%%%%%%%%%%%%%%%%%%%%%%%%%%%%%%%%

The language of \emph{epistemic specifications}  (or epistemic logic programs), developed by~\citeANP{gelfond91a} in three consecutive papers~\cite{gelfond91a,GelfondP93,gelfond94},
is an extension of disjunctive logic programming that introduces modal constructs to quantify over the set of stable models~\cite{gellif88b} of a program.
These new constructs, called \emph{subjective literals}, have the form~$\K l$ and~$\M l$ where~$l$ is an \emph{objective literal}~$l$, that is, any atom $p$, its explicit negation $\sneg p$, or any of these preceded by default negation.
Intuitively, $\K l$ and~$\M l$ respectively mean that $l$ is true in every stable model (cautious consequence) or in some stable model (brave consequence) of the program.
In many cases, these subjective literals can be seen as simple queries, but what makes them really interesting is their use in rule bodies,
which may obviously affect the set of stable models they are quantifying.
This feature makes them suitable for modelling introspection but, at the same time, may involve cyclic programs whose intuitive behaviour is not always easy to define.
In general, the semantics of an epistemic logic program may yield alternative sets of stable models, each set being called a \emph{world view}.
For instance, the epistemic program
\begin{eqnarray}
p \ \leftarrow \ \Not \, \K q \hspace{80pt}
q \ \leftarrow \ \Not \, \K p  \label{f:epiloop}
\end{eqnarray}
yields two world views $\{\{p\}\}$ and $\{\{q\}\}$, each one with a single stable model.
Deciding the intuitive world views of a cyclic program has motivated a wide debate in the literature.
This was mostly due to the fact that Gelfond's original \mbox{semantics (G91)} manifests a kind of self-supportedness or \emph{unfoundedness} typically illustrated by the epistemic program
\begin{eqnarray}
p \ \leftarrow \ \K p \label{f:selfsupport}
\end{eqnarray}
whose G91 world views are $\{\emptyset\}$ (as expected) but also $\{\{p\}\}$, which seems counterintuitive.
Other semantics~\cite{kawabagezh15,faheir15a,sheeit17a} managed to deal with this and other examples but fail to satisfy the elementary splitting property presented in~\cite{cafafa19a}, something that was preserved by the original G91.
Moreover, a first formalisation of foundedness was provided in~\cite{cafafa19b} and all the previously existing semantics violated that condition, except the new approach presented in that paper, FAEEL, which corresponds to a strengthening of G91 plus an extra foundedness check.
Thus, to the best of our knowledge, FAEEL is the only semantics satisfying both splitting and foundedness up to date.

There exist several implemented solvers for epistemic logic programs -- see~\cite{leckah18} for a recent survey.
Although there is no solver for FAEEL yet, the closest existing tools are those based on G91, since both semantics coincide in all epistemic logic programs whose subjective literals do not form positive cycles~\cite{fandinno19}.
This suggests that a solver for FAEEL can be constructed by applying an extra foundedness check on top of a G91 solver.
In fact, so far, all practical scenarios existing in the literature that involve epistemic problems can be represented without positive subjective cycles, so their computation in terms of G91 is sound with respect to FAEEL too.

In this paper, we present the \eclingo{} system\footnote{\url{https://github.com/potassco/eclingo}}, a solver for epistemic logic programs under the G91 semantics.
The tool is built on top of the ASP solver \clingo~\cite{gekakasc17a}
making use of its features for syntactic extensions and multi-shot solving.
The basic strategy applied by \eclingo{} is a guess-and-check method where the truth value of subjective literals is first guessed with choice rules for auxiliary atoms and, in a second step, the obtained values for those atoms are checked using the sets of cautious and brave consequences of the program.
This basic strategy has been improved with several optimisations.
We have made experiments on several scenarios for a couple of benchmark domains and compared \eclingo\ to
\wviews~\cite{Kelly07} (another solver for G91) and \EPASP~\cite{SLKL17},
which computes a close semantics~\cite{kawabagezh15} also accepted by \eclingo{},
and show that \eclingo\ outperforms these tools in most cases.

The rest of the paper is organised as follows. 
In the next section, we recall the basic definition of G91 semantics for epistemic logic programs.
In Section~\ref{sec:input}, we explain the input language of \eclingo{} and illustrate its usage with an example.
The next two sections respectively explain the basic process and some implemented optimisations.
Section~\ref{sec:eval} contains a comparison to solvers \wviews{} and \EPASP{} on a pair of benchmark domains and, finally, Section~\ref{sec:conc} concludes the paper.
%\red The rest of the paper is organised as follows. \dots \black

%%%%%%%%%%%%%%%%%%%%%%%%%%%%%%%%%%%%%%%%%%%%%%%%%%%%%%%%%%%%%%%%%%%%%%%%%%%%%%%
\section{Background}
\label{sec:background}
%%%%%%%%%%%%%%%%%%%%%%%%%%%%%%%%%%%%%%%%%%%%%%%%%%%%%%%%%%%%%%%%%%%%%%%%%%%%%%%
We assume some familiarity with the answer set semantics~\cite{GL91} for logic programs.
Given a set of atoms $\At$, an \emph{objective literal} is either an atom, a truth constant%
\footnote{For a simpler description of program transformations, we allow truth constants where $\top$ denotes true and $\bot$ denotes false.
  These constants can be easily removed.},
that is, \mbox{$a \in \At \cup \{\top,\bot\}$}, or an atom preceded by one or two occurrences of default negation, $\Not$.
We assume that, for each atom $a \in \At$, we have another atom `$\sneg a$' in \At\ that stands for the explicit negation of $a$.
As usual, the answer sets of a standard program $\Pi$, denoted as $\AS[\Pi]$, are those stable models of $\Pi$ that do not contain both $a$ and $\sneg a$.
The syntax of epistemic logic programs is an extension of ASP.
An expression of the form $\K l$, $\M l$ with $l$ being an objective literal, is called \emph{subjective atom}.
A \emph{subjective literal} can be a subjective atom $A$ or its default negation $\Not A$.
A \emph{rule} is an expression of the form:
\begin{gather}
a_1 \vee \dots \vee a_n \ \leftarrow \ L_1, \dots, L_m
	\label{eq:rule}
\end{gather}
with $n\geq 0$ and $m\geq 0$, where each $a_i \in \At$ is an atom and each $L_j$ a literal, either objective or subjective.
As usual, the left and right hand sides of~\eqref{eq:rule} are respectively called the \emph{head} and \emph{body} of the rule.
An \emph{epistemic program} (or epistemic specification) is a set of rules.
Given an epistemic program $\Pi$, we define $\At(\Pi)$ as the set of all atoms that occur in program~$\Pi$.
Similarly, by~$\Heads(\Pi)$, we the denote the set of all atoms that occur in the head of any rule in~$\Pi$ and, by $\Facts(\Pi)$, we denote the set of atoms that occur as facts in~$\Pi$.
Note that~$\Facts(\Pi) \subseteq \Head(\Pi) \subseteq \At(\Pi)$.
Let $\wv$ be a set of interpretations.
We write $\wv \models \bL \ell$ if objective literal $\ell$ holds (under the usual meaning) in all interpretations of $\wv$ and $\wv \models \Not \bL \ell$ if $\ell$ does not hold in some interpretation of~$\wv$.

\begin{definition}[Subjective reduct]
The \emph{subjective reduct} of an epistemic program $\Pi$ with respect to a set of propositional interpretations~$\wv$, written $\Pi^\wv$, is
obtained by replacing each subjective literal $L$ by $\top$ if $\wv \models L$ and by $\bot$ otherwise.\qed
\end{definition}

Note that the subjective reduct of an epistemic program does not contain subjective literals, and so, it is a standard logic program.
Therefore, we can collect its answer sets $\AS[\Pi^\wv]$.
We say that a set of propositional interpretations $\wv$ is a \emph{world view} of an epistemic program $\Pi$ if $\wv=\AS[\Pi^\wv]$.
Early works on epistemic specifications allowed for empty world views $W=\emptyset$ when the program has no answer sets, rather than leaving the program without world views.
Since this feature is not really essential, we exclusively refer to non-empty world views in this paper.
The complexity for deciding whether an epistemic program has a world view is $\Sigma^{P}_{3}$~\cite{truszczynski11b}, that is,
one level higher in the polynomial hierarchy than the complexity of (disjunctive) ASP, which is $\Sigma^{P}_{2}$.

We conclude this section by introducing a well-known example from~\cite{gelfond91a}.

\begin{example}\label{ex:college}
A given college uses the following set of rules to decide whether a student $X$ is eligible for a scholarship:
\begin{eqnarray}
\eligible(X) & \leftarrow & \high(X)   \label{ex:college.1} \\
\eligible(X) & \leftarrow & \minority(X),\, \fair(X) \label{ex:college.2}\\
\sneg\eligible(X) & \leftarrow & \sneg\fair(X),\, \sneg\high(X) \label{ex:college.3}\\
\interview(X) &\leftarrow  &\Not \K \eligible(X),\,
\Not \K \sneg \eligible(X) \label{ex:college.4}
\end{eqnarray}
Here, $\high(X)$ and $\fair(X)$ refer to the grades of student $X$.
The epistemic rule \eqref{ex:college.4} is encoding the college criterion
``\emph{The students whose eligibility is not determined by the college rules should be interviewed by the scholarship committee}.''
\qed
\end{example}
For instance, if the only available information for some student $mike$ is the disjunction
\begin{gather}
\fair(mike) \vee \high(mike) \label{ex:college.5}
\end{gather}
then the epistemic program \eqref{ex:college.1}-\eqref{ex:college.5} has a unique world view whose answer sets are:
\begin{eqnarray}
&\{\fair(mike),\,\interview(mike)\}& \label{f:sm1}\\
&\{\high(mike),\,\eligible(mike),\,\interview(mike)\} &\label{f:sm2}
\end{eqnarray}

%%%%%%%%%%%%%%%%%%%%%%%%%%%%%%%%%%%%%%%%%%%%%%%%%%%%%%%%%%%%%%%%%%%%%%%%%%%%%%%
\section{Using \texttt{eclingo}}
\label{sec:input}

As said before, \eclingo{} is based on \clingo's facilities (through its Python API) for syntax extension and multi-shot solving.
As a result, \eclingo's input language is just a minor extension of the input language accepted by \gringo~\cite{gekaosscth09a},
the grounder used by \clingo.
In this way, \eclingo{} programs can be constructed with three different types of statements: \emph{rules}, \emph{show} statements and \emph{constant} definitions.

The structure of a \clingo\ (or \eclingo{}) rule is as follows:
\begin{align*}
    H_1,\dots,H_m  \text{ \texttt{:-} }  B_1,\dots,B_n.
\end{align*}
where the head is formed by \clingo\ literals $H_i$ and
the body consists of elements $B_i$ that can be \clingo{} literals or subjective literals.
As happens in sequent calculus, commas in the antecedent (the body) represent a conjunction whereas commas in the consequent (the head) represent a disjunction.
The notation for subjective literals in \eclingo{} is as follows.
An expression $\K l$ is represented as a \clingo\ theory atom {\tt \&k\{$l$\}} where $l$ is a regular, objective literal, that is, it may combine an atom with default or explicit negation.
The only particularity is that default negation $\Not$ inside a {\tt \&k} operator must be replaced by the symbol $\sim$ (this limitation will be changed in the future).
For instance, the subjective literal $\K \Not\, \sneg p$ is currently represented in \eclingo{} as \mbox{\tt \&k\{$\sim$\,\sneg p\}}.
Operator $\M$ is not directly supported but any literal $\M l$ can be represented as {\tt not \&k\{$\sim l$\}}.

For instance, program~\eqref{f:epiloop} is represented as the \eclingo{} file {\tt program1.lp}:
\begin{Verbatim}[frame=single]
p :- not &k{q}.
q :- not &k{p}.
\end{Verbatim}
To obtain all world views of the program we just make the call
\begin{verbatim}
eclingo -n 0 program1.lp
\end{verbatim}
getting the result:
\begin{Verbatim}[frame=single]
eclingo version 0.2.0
Solving...
Answer: 1
&k{ p }
Answer: 2
&k{ q }
SATISFIABLE
\end{Verbatim}
Each answer provided by \eclingo{} corresponds to some world view of the epistemic program $\Pi$, but expressed as the set $X$ of those subjective literals that hold in the world view.
The set $X$ is enough to determine the answer sets in the world view.
To see why, we can define the (non-epistemic) program $\Pi_X$ where all subjective literals are replaced by their truth value $\top$ or $\bot$ with respect to $X$.
The world view then just consists of the answer sets of $\Pi_X$.
For instance, answer 1, makes {\tt \&k\{p\}} true and {\tt \&k\{q\}} false and, under that assumption, {\tt program1.lp} produces the unique answer set {\tt \{p\}}.
We plan to include a future option to expand one or all world views into their sets of answer sets.

As with regular atoms in \clingo,
the input language of  \eclingo{} provides a \mbox{\tt \#show p/n} directive to select those subjective atoms that we want to be displayed in each world view.
The syntax for this directive is the same as in \clingo{}, where {\tt p} is the name of some predicate (or its explicit negation) and {\tt n} its arity, that is, the number of arguments.
The difference in \eclingo{} is that this directive refers to the predicates used in subjective atoms to be displayed in each world view.
For instance, if we add the line
\begin{Verbatim}[frame=single]
#show p/0.
\end{Verbatim}
to our previous example, the information for the second world view would just be empty, since we assume we only want to display subjective literals of the form {\tt \&k\{p\}}.

As a second example, the program consisting of \eqref{ex:college.1}-\eqref{ex:college.5} from Example~\ref{ex:college} is represented in \eclingo{} as:
\begin{Verbatim}[frame=single]
eligible(X) :- high(X).
eligible(X) :- minority(X), fair(X).
-eligible(X) :- -fair(X), -high(X).
interview(X) :- not &k{ eligible(X)}, not &k{ -eligible(X)}, student(X).
student(mike).
fair(mike),high(mike).
\end{Verbatim}
where we have just introduced predicate {\tt student} to make variable {\tt X} safe in the epistemic rule for {\tt interview}.
The unique world view obtained by \eclingo{} in this example shows an empty list of subjective atoms meaning that both \mbox{\tt \&k\{eligible(mike)\}} and \mbox{\tt \&k\{-eligible(mike)\}} are false.
If we want to know what happens to predicate {\tt interview} we can add the line:
\begin{Verbatim}[frame=single]
#show interview/1.
\end{Verbatim}
to obtain now the output
\begin{Verbatim}[frame=single]
Solving...
Answer: 1
&k{ interview(mike) }
SATISFIABLE
\end{Verbatim}
%\subsubsection{Constant definitions}
%eclingo delegates the processing of constant definitions to \clingo. Thus, it does not provide any syntactical or semantical modification. Both \clingo and \eclingo{} define constants under the following structure:
%\begin{align*}
%    \texttt{\#const \emph{name}=\emph{value}.}
%\end{align*}
%where \texttt{\emph{name}} is the name of the constant, and \texttt{\emph{value}} is its value. For example, to specify a constant \texttt{length} with value \texttt{2} we write:
%\begin{align*}
%    \texttt{\#const length=2.}
%\end{align*}

\section{Basic solving process}
\label{sec:impl}
As we have seen, \eclingo's input language is a minor modification of the one for \clingo{}.
This is possible thanks to the parsing methods of \clingo's API for obtaining the representation of an epistemic program as an abstract syntax tree (AST)
in the form of a Python object.
In that way, program transformations can be easily combined with the usual \clingo{} functionality.
The \eclingo\ main algorithm solves epistemic logic programs following a guess and check strategy.
In the guessing phase, subjective literals are replaced by auxiliary atoms and a regular logic program is generated.
In the case of G91 semantics, this replacement of subjective literals is as follows.
For each objective literal~$\ell$, we define its corresponding auxiliary atom $\mathit{aux}_\ell$ as:
\[
\mathit{aux}_\ell \eqdef \left\{
\begin{array}{ll}
\mathit{aux}\_p & \text{if } \ell=(p) \\
\mathit{aux\_not}\_p & \text{if } \ell=(\sim p) \\
\mathit{aux\_sn}\_p & \text{if } \ell=(- p) \\
\mathit{aux\_not\_sn}\_p & \text{if } \ell=(\sim - p)
\end{array}
\right.
\]
for any atom $p$. 
Now, each positive subjective literal $\mathtt{\&k\{}\ell \mathtt{\}}$ in the epistemic program is replaced by $(\texttt{not\ not\ } \mathit{aux}_\ell)$ whereas each negated subjective literal $\mathtt{not \ \&k\{}\ell \mathtt{\}}$ is just replaced by $(\texttt{not\ } \mathit{aux}_\ell)$.
Additionally, for each auxiliary atom $\mathit{aux_\ell}$, we add the choice rule:
\begin{align*}
    \mathtt{\{}\ aux_\ell \ \mathtt{\}.}
\end{align*}
The result of this translation is a regular logic program that can now be used for guessing the truth values of subjective literals, represented as auxiliary atoms.
Thus, when asking \clingo\ to solve this program using its multi-shot feature, it will go returning answer sets that constitute potential \emph{candidates} for world views.
Since we only retrieve the auxiliary atoms $\mathit{aux\_x}$ from each answer set, we may have repeated answers (due to differences in the rest of atoms that are hidden).
For this reason, we use the \clingo{} ``projection'' option to rule out these duplicates.

%\subsection{Solving an epistemic logic program} \label{subsec:solve}
%In the first one, \eclingo{} generates a set of possible solutions to the original program, feeding \clingo\ with the auxiliary program and filtering
%the output answer sets.
%The world view representatives, that is, the auxiliary subjective atoms, are isolated and stored as potential solutions.
%\sidecomment{One at a time?}

In the checking phase, \eclingo{} verifies that each candidate actually constitutes a valid world view.
To this aim, we need to check several conditions on the subjective literals with respect to the answer sets of the candidate world view:
\begin{enumerate}
    \item For each subjective literal \texttt{\&k\{$\ell$\}}, literal $\ell$ must be in every answer set.

    \item For each subjective literal \texttt{not \&k\{$\ell$\}}, literal $\ell$ cannot be in every answer set.

    \item For each subjective literal \texttt{\&k\{$\sim \ell$\}}, literal $\ell$ cannot be in any answer set.

    \item For each subjective literal \texttt{not \&k\{$\sim \ell$\}}, literal $\ell$ must be in some answer set.
\end{enumerate}
To obtain the answer sets of a candidate world view $X$, we would have to expand all the answers for $\Pi_X$ provided by \clingo{}. 
Fortunately, this expansion can be avoided since the four conditions above can be checked using \clingo{} modes for \emph{cautious} and \emph{brave} reasoning, that are computed by iterated intersection and union operations, respectively.
Let $\mathit{cautious}(\Pi_X)$ and $\mathit{brave}(\Pi_X)$ denote the set of atoms in the cautious and brave consequences of $\Pi_X$, respectively.
%, which rely on basic set operations over the resulting answer sets: intersection and union, respectively.
In particular, we can reduce those four conditions to:
\begin{enumerate}
    \item For each subjective literal \texttt{\&k\{$\ell$\}}, check $\ell \in \mathit{cautious}(\Pi_X)$.

    \item For each subjective literal \texttt{not \&k\{$\ell$\}}, check $\ell \not\in \mathit{cautious}(\Pi_X)$.

    \item For each subjective literal \texttt{\&k\{$\sim \ell$\}}, check $\ell \not\in \mathit{brave}(\Pi_X)$.

    \item For each subjective literal \texttt{not \&k\{$\sim \ell$\}}, check $\ell \in \mathit{brave}(\Pi_X)$.
\end{enumerate}

Although \eclingo{} was proposed as a solver for epistemic specifications under G91 semantics, it also supports
the semantics proposed in~\cite{kawabagezh15} (K15), which can be obtained by a simple transformation.
In particular, K15 can be obtained from G91 by replacing each expression $\bL \ell$ in the original epistemic program by the conjunction $\bL \ell \wedge \ell$.
When $\bL \ell$ is not preceded by $\Not$, this simply generates an additional objective literal $\ell$ in the rule body.
However, when we have to replace $\Not \bL \ell$ by $\Not (\bL \ell \wedge \ell)$, we obtain the disjunction $\Not \bL \ell \vee \Not \ell$ that is replaced by a new auxiliary atom $\mathit{aux}$ that is, then, defined by the pair of rules:
\begin{eqnarray*}
\mathit{aux} & \leftarrow & \Not \bL \ell \\
\mathit{aux} & \leftarrow & \Not \ell
\end{eqnarray*}
Then, the rest of the translation proceeds as with G91.

\section{Optimising the solving process}
\label{sec:optim}
Several optimisations have been implemented on top of the basic solving process presented above.
%The direct implementation of the solving method presented before is a supposes a naive approach to solve epistemic logic programs. As we already said, \eclingo{} pretends to meet a high level of efficiency. For this purpose, we propose a set of optimizations or heuristics for this method.
%
A first optimisation is the addition of \emph{consistency constraints}.
Notice that, once a subjective atom {\tt \&k\{$\ell$\}} is replaced by a standard auxiliary atom $\mathit{aux}$, the relation to the original literal is lost.
Thus, the guess may produce epistemically inconsistent combinations like, for instance, an answer set containing:
\begin{align*}
	\texttt{\{\&k\{$\sim$\,p\}, p\}}
\end{align*}
%The subjective literal means that \texttt{p} is false in every answer set while in this proper one \texttt{p} is true. The same would happen in the opposite situation:
%\begin{align*}
%	\texttt{\{\&k\{p\}\}}
%\end{align*}
%where \texttt{p} must be true in every answer set while in this case, it is false.
%This situation can lead to a bigger set of possible candidates to check, where each check is resolved, as we saw, by a $\Sigma^{P}_{2}$ routine (i.e., an invocation to \clingo).
This problem can be avoided by adding the rule:
\begin{align*}
&\text{\texttt{:-}} \; \mathit{aux}_\ell,\, \overline{\ell}.
\end{align*}
%As we can see, rule \eqref{r:1} means that if a literal has to be true in every answer set it also has to be in this one itself. In the same way, rule \ref{r:2} means that if a literal has to be false in every answer set it also has to be in this one itself.
%
for each subjective literal of the form {\tt \&k\{$\ell$\}}, where $\overline{\ell} \eqdef \Not \ell$ if $\ell$ does not contain $\sim$ and $\overline{\sim\ \alpha}=\alpha$ if $\ell = (\sim \alpha)$.
These constraints improve the efficiency of \eclingo{} by ruling out epistemically inconsistent world views during the guessing phase, without requiring their subsequent check.

Another optimisation implemented in \eclingo{} consists in using the grounder {\tt gringo} to approximate the well-founded model (WFM) of the auxiliary guess program $\Pi$.
Computing the WFM of a logic program takes a polynomial complexity in the general case, while computing the stable models of $\Pi$ is $\Sigma^{P}_{2}$. This difference makes this heuristic a worthwhile strategy. 
The WFM of a program $\Pi$ is a three-valued interpretation we can describe as a pair of disjoint sets of atoms $\tuple{I^+,I^-}$ respectively collecting the true and false atoms in the model, being all the rest undefined.
It is well-known that $I^+ \subseteq \mathit{cautious}(\Pi)$ and $I^- \subseteq \at \setminus \mathit{brave}(\Pi)$.
When {\tt gringo} processes any program $\Pi$ (even if it is originally ground) it performs some simplifications that allow retrieving an approximation of the WFM $\tuple{I^+,I^-}$ of $\Pi$.
In particular, if $\Pi'$ is the result provided by {\tt gringo}, then $\Facts(\Pi') \subseteq I^+$ and $\at \setminus \Heads(\Pi') \subseteq I^-$ which, in their turn, imply $\Facts(\Pi') \subseteq \mathit{cautious}(\Pi)$ and $\Heads(\Pi') \subseteq \mathit{brave}(\Pi)$.
%
%semantics is a well-known theory for negation as failure. Roughly, computing the well-founded model of a logic programming consists of partitioning the set of ground atoms into three sets that define its truth value: true, false, and unknown. This partitioning is performed by focusing on the structure of the rules. In this way, an atom is true if it appears alone in the head of a rule and that rule has no body (i.e., it is a fact). Similarly, an atom that does not appear in the head of any rule is considered false, since it is no way to derive it. Finally, the rest of the atoms (i.e., those that cannot be stated as true or false) are categorized as unknown.
%
%The proper definition of the well-founded model of a logic program allows to state some relations between the set of stable models of the logic program and the well-founded model itself. For example:
%\begin{theorem}
%An atom considered true in the well-founded model must belong to every stable model of the program.
%\end{theorem}
%\begin{theorem}
%An atom considered false in the well-founded model must not belong to any stable model of the program.
%\end{theorem}
%
%As can be seen, these properties are easily relatable with the epistemic semantics:
%\begin{theorem}
%The subjective atom Kp holds for every p true in the well-founded model of a logic program.
%\end{theorem}
%\begin{theorem}
%The subjective atom K \textapprox p holds for every p false in the well-founded model of a logic program.
%\end{theorem}
As a result, if we only use the grounder {\tt gringo} on the guess program $\Pi$ to obtain $\Pi'$ we get a good estimate of regular atoms that are always true (resp.\ always false) in all the answer sets of the program.
In particular, if we get $p \in \Facts(\Pi')$ is true, then $\K p$ holds and we can safely add the corresponding auxiliary atom {\tt aux\_p}.
The same happens if atom $p \not\in \Heads(\Pi')$ we can conclude $\K \Not \, p$ and add the auxiliary atom {\tt aux\_not\_p}.
These true subjective literals are added to the guess program $\Pi$ to reduce the search space before computing the answer sets.
Of course, this behaviour is implemented in an iterative way until no new addition is made, as described in Algorithm~\ref{alg:wfm}.
Here, we use the set $K^+$ (resp. $K^-$) to collect every atom {\tt p} occurring in some expression $\mathtt{\&k\{p\}}$ (resp. $\mathtt{\&k\{\sim p\}}$) in the original epistemic program.
The algorithm uses two set variables, one for facts $F$ and one for heads $H$, that are initially set to $\emptyset$ and $\at(\Pi)$, respectively.
Then, we repeat calls to {\tt gringo}'s function ground($\Pi$) while we obtain some new fact {\tt p} originally used in a subjective literal $\mathtt{\&k\{p\}}$ or we lose some head atom {\tt p} originally used in a subjective literal $\mathtt{\&k\{\sim p\}}$.
When this happens, we include the corresponding auxiliary atoms in the program.
\SetKwRepeat{Do}{do}{while}%
\begin{algorithm}[Htbp]
\SetAlgoLined
$F := \emptyset$; \ $H := \at(\Pi)$;\\
$\Pi$ := ground($\Pi$);\\
\While{$(\Facts(\Pi)\setminus F)\cap K^+ \neq \emptyset$ {\bf or} $(H \setminus \Heads(\Pi)) \cap K^- \neq \emptyset$}{
	\ForEach{$p \in (\Facts(\Pi)\setminus F) \cap K^+$}{
	    $\Pi$ := $\Pi \; \cup \; \{ \mathtt{aux\_p} \}$;\\
	}
	\ForEach{$p \in (H \setminus \Heads(\Pi)) \cap K^-$}{
	    $\Pi$ := $\Pi \; \cup \; \{ \mathtt{aux\_not\_p} \}$;\\
	}
	$F:= \Facts(\Pi)$; \ $H:= \Heads(\Pi)$; \\
    $\Pi$ := ground($\Pi$);
}
\caption{Extending an epistemic logic program using {\tt gringo} grounder.}
\label{alg:wfm}
\end{algorithm}
To illustrate the algorithm, take Fig.~\ref{fig:p1} showing an \eclingo{} input program on the left and its corresponding guess program $\Pi_1$ on the right.
\begin{figure}[htbp]
\begin{subfigure}{.4\textwidth}
\begin{lstlisting}[frame = single, mathescape, basicstyle = \ttfamily, belowskip=37pt]
a :- not b.
c :- &k{$\sim$a}.
d :- not &k{$\sim$e}.
p :- &k{$\sim$d}.
\end{lstlisting}
\end{subfigure}
\hspace{25pt}
\begin{subfigure}{.4\textwidth}
\begin{lstlisting}[frame = single, mathescape, basicstyle = \ttfamily]
a :- not b.
c :- aux_a.
d :- not aux_not_e.
p :- aux_not_d.
{aux_a}.
{aux_not_e}.
{aux_not_d}.  
\end{lstlisting}
\end{subfigure}
\setlength{\abovecaptionskip}{5pt}
\caption{An \eclingo{} program (left) and its corresponding guess program $\Pi_1$ (right).}
\label{fig:p1}
\end{figure}
The latter generates 8 possible candidate world views corresponding to all the free combinations of truth values for the auxiliary atoms.
However, if we run {\tt gringo} on $\Pi_1$ we obtain the new ground program $\Pi_2$:
\begin{Verbatim}[frame=single]
a.
c :- aux_a.
d :- not aux_not_e.
p :- aux_not_d.
{aux_a}.
{aux_not_e}.
{aux_not_d}.  
\end{Verbatim}
where $\Facts(\Pi_2)=\{a\}$ and $\Heads(\Pi_2)$ consists of $\{a,c,d,p\}$ and the auxiliary atoms.
As a result, we know that $\bL a$ and $\bL \Not\; e$ must hold, and so, atoms {\tt aux\_a} and {\tt aux\_not\_e} can be added to $\Pi_2$ or just replaced by $\top$ ({\tt \&true} in {\tt gringo} notation).
After doing that replacement on $\Pi_2$ if we run {\tt gringo} again we obtain $\Pi_3$:
\begin{Verbatim}[frame=single]
a.
c.
p :- aux_not_d.
{aux_not_d}.
\end{Verbatim}
but, as we can see, $d \not\in \Heads(\Pi_3)$ and we conclude $\bL \Not \; d$ so atom {\tt aux\_not\_d} can be replaced by $\top$.
The resulting program has no auxiliary predicates and no new changes will occur after grounding, so the algorithm stops.
In this example, the optimisation has solved the problem even before the guessing phase.
This is because subjective literals where stratified in the original program.
In the general case, however, the {\tt gringo}-based optimisation is not so efficient if we have cyclic dependencies among the subjective literals.

%%%%%%%%%%%%%%%%%%%%%%%%%%%%%%%%%%%%%%%%%%%%%%%%%%%%%%%%%%%%%%%%%%%%%%%%%%%%%%%
\section{Evaluation and comparison to other solvers}
\label{sec:eval}
%%%%%%%%%%%%%%%%%%%%%%%%%%%%%%%%%%%%%%%%%%%%%%%%%%%%%%%%%%%%%%%%%%%%%%%%%%%%%%%

In this section,
we compare \eclingo{} with other two epistemic solvers, \wviews~\cite{Kelly07} and \EPASP~\cite{SLKL17}, both with respect to usability and efficiency.
%
%\subsection{Input Language}
%\subsubsection{\wviews}
The tool \wviews\footnote{\url{https://github.com/galactose/wviews}} is a solver for epistemic specifications under G91 semantics developed by Michael Kelly for his Honour's Thesis. It is built upon the ASP system {\tt DLV} and allows the inclusion of subjective literals under a  simple notation. However, this simplicity is eclipsed by the limitations in the rest of its grammar. \wviews{} parser is very sensitive to minor changes in the problem representation. In fact, we experienced problems to execute \wviews{} on programs with predicates with more than one argument, something that forced us to test the benchmarks for planning on ground programs.
A peculiarity of \wviews{} is that it computes \emph{all} the world views of an epistemic program.

%\subsubsection{\EPASP}
\EPASP\footnote{\url{https://github.com/tiep/EP-ASP}}~\cite{SLKL17} is another solver for epistemic logic programs that can compute world views under two semantics: \cite{kawabagezh15} (K15)  (also computed by \eclingo{}) and~\cite{sheeit17a}. Just like \eclingo{}, it is built upon the ASP system \clingo, but using version 4.5.3. \EPASP{} input programs are generated using another independent and non-integrated tool, {\tt ELPS}~\cite{BK14}, that provides a method to translate an epistemic logic program with sort definitions into a standard ASP program. The grammar that defines a correct input program for {\tt ELPS} is, therefore, substantially different from the one used by \clingo{}. It considers four types of statements: directives, sort definitions, predicate declarations and rules. Directives can be either a constant definition or a {\tt maxint} declaration, so the range for numerical expressions is predefined, unlike in \clingo{}.
%Regarding the rest of the program, subjective literals can be included in the body of the rules under a pretty simple syntax.

%the need of the independent and non-integrated tool {\tt ELPS} can be seen as a disadvantage, diusing two different, not fully integrated programs for solving the epistemic logic program can be seen as a difficulty in its usage. Also, while this epistemic logic program represents subjective literals in a very comfortable way, the way it is split makes the representation of the rest of the program a hard task.

Regarding the efficiency comparison, we have executed the three tools (i.e., \wviews, \EPASP{} and \eclingo) on scenarios for two well-known problems in the literature: the Eligibility problem (Example~\ref{ex:college}) and a variant of the Yale Shooting problem with incomplete knowledge about the initial state, looking for a conformant plan (that is, a plan that always succeeds, regardless of the initial state).
Experiments were performed on a machine equipped with an Intel i7-8550U (up to 4.0GHz) and 8GB memory running Ubuntu 18.04.4 LTS.
% Experiments were performed using a set of \texttt{Amazon Web Services}\footnote{\url{https://aws.amazon.com/}} virtual machines. These virtual machines are instances of type \texttt{t2.micro} using 1 vCPU with processor {\tt Intel Xeon} family, clock speed 2.5 GHz and 1 GiB of memory.
% The chosen operating system is an Ubuntu Server 18.04 LTS, given as an AMI~\footnote{Ubuntu Server 18.04 LTS (HVM) 64-Bit(x86). AMI ID: \texttt{ami-085925f297f89fce1}}, this is, an \emph{Amazon Machine Image}.
%
The times were measured using a Python wrapper and taking the average of 10 different executions for each problem instance.
%
% To avoid irregularities due to the server performance, a delay of 120s between each two executions was introduced.
%
%The selected problems for the benchmarking are the Scholarship Eligibility Problem, since it is the first problem upon epistemic specifications were proposed, and the epistemic version of the Yale Shooting Problem, to see how the tools face a harder problem as conformant planning is.
%
Encodings for \eclingo{} scenarios can be downloaded from the Git repository.
It is important to note that \EPASP{} encodings are already preprocessed by {\tt ELPS}, so our execution times do not consider this translation step.

Table \ref{tab:eligible} shows the average times for the Scholarship Eligibility Problem obtained by \wviews{} under G91 semantics, \EPASP{} under K15 semantics and \eclingo{} under both G91 and K15 semantics, to make a fair comparison. 
In this problem, the tools were fed with 25 scenarios denoted as eligibleXX were XX is the number of the instance and, at the same time, the number of students for the problem. 
As can be seen, \wviews{} can only solve the first 8 scenarios (with a timeout of 2 minutes) and only performs better than \eclingo{} in the first one. 
The performance of \eclingo{} is more robust, solving 21 instances clearly below 1 second, and showing a slight grow (up to 3.35s) for the 4 larger instances.
Note that \eclingo{} is computing all the world views of the epistemic program, since this is default mode for \wviews{}.
However, in the comparison with \EPASP{} (the last three columns) we just look for one world view.
The options we used for that solver were {\tt  pre=1, max=0} (brave/cautious preprocessing, and K15 semantics, respectively).
\EPASP{} solves 16 scenarios and reaches the timeout of 120s for the 9 remaining ones. For the first 9 scenarios (with the exception of {\tt eligible06}), \EPASP{} execution times are better but very close to \eclingo{} ones. However, in the examples from 10 to 16 the performance of \EPASP{} is clearly worse and, moreover, shows an unpredictable variability among the solved cases, from 0.063s in {\tt eligible15} and {\tt eligible16} to 44.96s for {\tt eligible12}. \eclingo{} under K15 solves the 25 scenarios in a range of times from 0.03s to 0.05s, except {\tt eligible25} that just takes 0.54s.
When using G91 semantics, we get the same the world views (for this domain) but the \eclingo{} times are slightly better, since K15 is computed as a translation to G91.

\begin{table}[ht]
\begin{tabular}{c|cc|ccc}
\cline{1-6}
  & \multicolumn{2}{c|}{Computing all world views} & \multicolumn{3}{c}{Computing one world view} \\ \cline{1-6}
  & \texttt{Wviews G91} & \texttt{eclingo G91} & \texttt{EP-ASP K15} & \texttt{eclingo K15} & \texttt{eclingo G91} \\ \cline{1-6}
\texttt{eligible01} & \textbf{0.025}     & 0.035   & \textbf{0.024}    & 0.033    & 0.034    \\
\texttt{eligible02} & 0.042   & \textbf{0.036}   & \textbf{0.021}    & 0.034    & 0.035           \\
\texttt{eligible03} & 0.103   & \textbf{0.035}   & \textbf{0.022}    & 0.035    & 0.033           \\
\texttt{eligible04} & 0.347   & \textbf{0.036}   & \textbf{0.023}    & 0.036    & 0.034           \\
\texttt{eligible05} & 1.397   & \textbf{0.036}   & \textbf{0.025}    & 0.035    & 0.034           \\
\texttt{eligible06} & 5.728   & \textbf{0.037}   & 0.138             & 0.037    & \textbf{0.035}  \\
\texttt{eligible07} & 27.271  & \textbf{0.037}   & \textbf{0.031}    & 0.037    & 0.036           \\
\texttt{eligible08} & 113.188 & \textbf{0.037}   & \textbf{0.032}    & 0.037    & 0.036           \\
\texttt{eligible09} & -       & \textbf{0.039}   & \textbf{0.035}    & 0.039    & 0.037           \\
\texttt{eligible10} & -       & \textbf{0.041}   & 1.795             & 0.039    & \textbf{0.037}  \\
\texttt{eligible11} & -       & \textbf{0.048}   & 12.302            & 0.041    & \textbf{0.040}  \\
\texttt{eligible12} & -       & \textbf{0.049}   & 44.959            & 0.043    & \textbf{0.040}  \\
\texttt{eligible13} & -       & \textbf{0.049}   & 0.934             & 0.044    & \textbf{0.041}  \\
\texttt{eligible14} & -       & \textbf{0.050}   & 13.574            & 0.045    & \textbf{0.041}  \\
\texttt{eligible15} & -       & \textbf{0.048}   & 0.063             & 0.044    & \textbf{0.042}  \\
\texttt{eligible16} & -       & \textbf{0.085}   & 0.063             & 0.054    & \textbf{0.040}  \\
\texttt{eligible17} & -       & \textbf{0.150}   & -                 & 0.041    & \textbf{0.040}  \\
\texttt{eligible18} & -       & \textbf{0.142}   & -                 & 0.043    & \textbf{0.041}  \\
\texttt{eligible19} & -       & \textbf{0.392}   & -                 & 0.050    & \textbf{0.042}  \\
\texttt{eligible20} & -       & \textbf{0.414}   & -                 & 0.049    & \textbf{0.041}  \\
\texttt{eligible21} & -       & \textbf{0.567}   & -                 & 0.049    & \textbf{0.042}  \\
\texttt{eligible22} & -       & \textbf{1.516}   & -                 & 0.049    & \textbf{0.043}  \\
\texttt{eligible23} & -       & \textbf{1.015}   & -                 & 0.051    & \textbf{0.044}  \\
\texttt{eligible24} & -       & \textbf{0.937}   & -                 & 0.050    & \textbf{0.044}  \\
\texttt{eligible25} & -       & \textbf{3.347}   & -                 & 0.544    & \textbf{0.048}  \\ \cline{1-6}
\end{tabular}
\caption{Eligibility Problem. Time in seconds: timeout fixed in 120s.}
\label{tab:eligible}
\end{table}

For the Yale shooting benchmark we actually extended the benchmarks from the \EPASP{} repository with the instances {\tt yale09} to {\tt yale13}, all for path length 10 and the last one without any world view, to try an unsatisfiable problem.
Table \ref{tab:yale} shows average execution times obtained by \EPASP{} under K15 semantics and \eclingo{} under G91 semantics. 
In this case, the comparison is less accurate than the eligibility example for several reasons. 
First, we have used two slightly different problem encodings. 
For \EPASP{}, we used K15 semantics on the benchmarks provided with the tool, already translated from {\tt ELPS}. Moreover, we used the recommended \EPASP{} configuration for planning and heuristics, passing the options {\tt pre=1, max=0, planning=1, heuristic=1}.
In this special configuration, \EPASP{} recognizes the action theory representation (fluents, actions, goals, etc) and is capable of applying planning based heuristics. In particular, \EPASP{} solves the problem first as a regular planning domain and then uses this result to prune the search space for the conformant planning problem. For \eclingo{}, we redesigned the epistemic rules in the scenario to use G91 instead, something that, under our understanding, provides a more natural use of the epistemic operators (see~\citeNP{cafafa19a} for a discussion). Besides, \eclingo{} does not apply any planning-based specific heuristics or optimisation: it treats all the scenarios as regular epistemic specifications\footnote{We also executed \EPASP{} for these benchmarks without using the planning mode, but it produced apparently incoherent results, immediately printing an empty world view.}.

As we can see in Table \ref{tab:yale}, \EPASP{} in planning mode performs slightly better than \eclingo{} in all the cases solved by both tools. However, \EPASP{} reaches the timeout in three scenarios for path length 10, while \eclingo{} solves all of them.
\begin{table}[ht]
    \begin{tabular}{c|cc|c}
\cline{1-4}
\multicolumn{4}{c}{Computing one world view}         \\ \cline{1-4}
           & \texttt{EP-ASP K15}, planning mode & \texttt{eclingo G91} & Path length \\ \cline{1-4}
\texttt{yale01}  & \textbf{0.025}    & 0.042       & 1           \\
\texttt{yale02}  & \textbf{0.024}    & 0.039       & 2           \\
\texttt{yale03}  & \textbf{0.027}    & 0.040       & 3           \\
\texttt{yale04}  & \textbf{0.027}    & 0.040       & 4           \\
\texttt{yale05}  & \textbf{0.033}    & 0.045       & 5           \\
\texttt{yale07}  & \textbf{0.042}    & 0.073       & 7           \\
\texttt{yale08}  & \textbf{0.036}   & 0.051       & 8           \\
\texttt{yale09}  & -                 & \textbf{0.314}       & 10          \\
\texttt{yale10}  & \textbf{0.068}   & 0.257       & 10          \\
\texttt{yale11}  & -                 & \textbf{0.106}       & 10          \\
\texttt{yale12}  & \textbf{0.119}    & 1.013       & 10          \\
\texttt{yale13}* & -                 & \textbf{0.445}       & 10          \\ \cline{1-4}
\end{tabular}
\setlength{\abovecaptionskip}{5pt}
\caption{Yale Shooting Problem. Time in seconds: timeout fixed in 120s. Problem yale13 is unsatisfiable.}
\label{tab:yale}
\end{table}

Due to lack of encodings and the difficulty shown by \wviews{} grammar to represent the epistemic Yale Shooting Problem, we generated a ground version of the problem. Thus, both \wviews{} and \eclingo{} are compared under this ground program. Table \ref{tab:ground_yale} shows the average times for 10 runs of 12 scenarios of the problem. As we can see, \wviews{} can only solve the first three scenarios while \eclingo{} can still solve all of them, although with worse execution times than \eclingo{} in the non-ground version (Table \ref{tab:yale}).
This is because the independent grounding we used is apparently less efficient, generating more ground subjective literals and creating harder instances.
%As in the Scholarship Eligibility Problem, \wviews{} does increase its execution time as the number of instance does. While \eclingo{} also increases its execution time progressively, \wviews{} exceeds the timeout at the 4th one.

\begin{table}[ht]
    \begin{tabular}{c|cc|c}
\cline{1-4}
\multicolumn{4}{c}{Computing all world views}             \\ \cline{1-4}
                   & \texttt{Wviews G91}   & \texttt{eclingo G91}  & Path length \\ \cline{1-4}
\texttt{ground\_yale01}  & 0.054     & \textbf{0.038}    & 1           \\
\texttt{ground\_yale02}  & 0.590     & \textbf{0.040}    & 2           \\
\texttt{ground\_yale03}  & 11.330    & \textbf{0.042}    & 3           \\
\texttt{ground\_yale04}  & -         & \textbf{0.046}    & 4           \\
\texttt{ground\_yale05}  & -         & \textbf{0.063}    & 5           \\
\texttt{ground\_yale07}  & -         & \textbf{0.230}    & 7           \\
\texttt{ground\_yale08}  & -         & \textbf{0.108}    & 8           \\
\texttt{ground\_yale09}  & -         & \textbf{2.137}    & 10          \\
\texttt{ground\_yale10}  & -         & \textbf{25.521}   & 10          \\
\texttt{ground\_yale11}  & -         & \textbf{28.702}   & 10          \\
\texttt{ground\_yale12}  & -         & \textbf{59.013}   & 10          \\
\texttt{ground\_yale13}* & -         & \textbf{2.278}    & 10          \\ \cline{1-4}
\end{tabular}
\setlength{\abovecaptionskip}{5pt}
\caption{Ground version of the Yale Shooting Problem. Time in seconds: timeout fixed in 120s. Problem ground\_yale13 is unsatisfiable.}
\label{tab:ground_yale}
\end{table}
%%%%%%%%%%%%%%%%%%%%%%%%%%%%%%%%%%%%%%%%%%%%%%%%%%%%%%%%%%%%%%%%%%%%%%%%%%%%%%%

\section{Conclusions and Related Work}
\label{sec:conc}
%%%%%%%%%%%%%%%%%%%%%%%%%%%%%%%%%%%%%%%%%%%%%%%%%%%%%%%%%%%%%%%%%%%%%%%%%%%%%%%

We have presented \eclingo{}, a solver for epistemic specifications under G91~semantics.
The solver is programmed on top of \clingo, using its syntactic extension and multi-shot solving features.
We have tested the execution of \eclingo{} and compared to other two epistemic solvers in a pair of domains from the literature.
The results seem to point out that \eclingo{} provides a better performance, especially in the number of solved scenarios.

Our future work includes the addition of other optimisation techniques and, more importantly, the implementation of an unfoundedness check to disregard self-supported world views that are sometimes produced by G91 semantics (although only on positive cycles), computing in this way the stronger semantics provided in~\cite{cafafa19b}.
We also plan to extend the benchmarks with harder instances that can be parameterised and possibly include comparisons to solvers under other semantics, on scenarios where it can be guaranteed that the same solutions are obtained.

\bibliographystyle{acmtrans}
\bibliography{krr,refs,procs}

\begin{thebibliography}{}

\bibitem[\protect\citeauthoryear{Balai and Kahl}{Balai and Kahl}{2014}]{BK14}
{\sc Balai, E.} {\sc and} {\sc Kahl, P.} 2014.
\newblock Epistemic logic programs with sorts.
\newblock In {\em Proceedings of the Workshop on Answer Set Programming and
  Other Computing Paradigms, 2014}, {D.~Inclezan} {and} {M.~Maratea}, Eds.

\bibitem[\protect\citeauthoryear{Balduccini, Lierler, and Woltran}{Balduccini
  et~al\mbox{.}}{2019}]{lpnmr19}
{\sc Balduccini, M.}, {\sc Lierler, Y.}, {\sc and} {\sc Woltran, S.}, Eds.
  2019.
\newblock {\em Proceedings of the Fifteenth International Conference on Logic
  Programming and Nonmonotonic Reasoning (LPNMR'19)}. Lecture Notes in
  Artificial Intelligence, vol. 11481. Springer-Verlag.

\bibitem[\protect\citeauthoryear{Cabalar, Fandinno, and {Fari{\~n}as del
  Cerro}}{Cabalar et~al\mbox{.}}{2019a}]{cafafa19b}
{\sc Cabalar, P.}, {\sc Fandinno, J.}, {\sc and} {\sc {Fari{\~n}as del Cerro},
  L.} 2019a.
\newblock Founded world views with autoepistemic equilibrium logic.
\newblock See \citeN{lpnmr19}, 134--147.

\bibitem[\protect\citeauthoryear{Cabalar, Fandinno, and {Fari{\~n}as del
  Cerro}}{Cabalar et~al\mbox{.}}{2019b}]{cafafa19a}
{\sc Cabalar, P.}, {\sc Fandinno, J.}, {\sc and} {\sc {Fari{\~n}as del Cerro},
  L.} 2019b.
\newblock Splitting epistemic logic programs.
\newblock See \citeN{lpnmr19}, 120--133.

\bibitem[\protect\citeauthoryear{Fandinno}{Fandinno}{2019}]{fandinno19}
{\sc Fandinno, J.} 2019.
\newblock Founded (auto)epistemic equilibrium logic satisfies epistemic
  splitting.
\newblock {\em Theory and Practice of Logic Programming\/}~{\em 19,\/}~5-6,
  671--687.

\bibitem[\protect\citeauthoryear{{Fari{\~n}as del Cerro}, Herzig, and {Iraz
  Su}}{{Fari{\~n}as del Cerro} et~al\mbox{.}}{2015}]{faheir15a}
{\sc {Fari{\~n}as del Cerro}, L.}, {\sc Herzig, A.}, {\sc and} {\sc {Iraz Su},
  E.} 2015.
\newblock Epistemic equilibrium logic.
\newblock In {\em Proceedings of the Twenty-fourth International Joint
  Conference on Artificial Intelligence (IJCAI'15)}, {Q.~Yang} {and}
  {M.~Wooldridge}, Eds. {AAAI} Press, 2964--2970.

\bibitem[\protect\citeauthoryear{Gebser, Kaminski, Kaufmann, and Schaub}{Gebser
  et~al\mbox{.}}{2019}]{gekakasc17a}
{\sc Gebser, M.}, {\sc Kaminski, R.}, {\sc Kaufmann, B.}, {\sc and} {\sc
  Schaub, T.} 2019.
\newblock Multi-shot {ASP} solving with clingo.
\newblock {\em Theory and Practice of Logic Programming\/}~{\em 19,\/}~1,
  27--82.

\bibitem[\protect\citeauthoryear{Gebser, Kaminski, Ostrowski, Schaub, and
  Thiele}{Gebser et~al\mbox{.}}{2009}]{gekaosscth09a}
{\sc Gebser, M.}, {\sc Kaminski, R.}, {\sc Ostrowski, M.}, {\sc Schaub, T.},
  {\sc and} {\sc Thiele, S.} 2009.
\newblock On the input language of {ASP} grounder gringo.
\newblock In {\em Proceedings of the Tenth International Conference on Logic
  Programming and Nonmonotonic Reasoning (LPNMR'09)}, {E.~Erdem}, {F.~Lin},
  {and} {T.~Schaub}, Eds. Lecture Notes in Artificial Intelligence, vol. 5753.
  Springer-Verlag, 502--508.

\bibitem[\protect\citeauthoryear{Gelfond}{Gelfond}{1991}]{gelfond91a}
{\sc Gelfond, M.} 1991.
\newblock Strong introspection.
\newblock In {\em Proceedings of the Ninth National Conference on Artificial
  Intelligence {(AAAI'91)}}, {T.~Dean} {and} {K.~McKeown}, Eds. {AAAI} Press /
  The {MIT} Press, 386--391.

\bibitem[\protect\citeauthoryear{Gelfond}{Gelfond}{1994}]{gelfond94}
{\sc Gelfond, M.} 1994.
\newblock Logic programming and reasoning with incomplete information.
\newblock {\em Annals of Mathematics and Artificial Intelligence\/}~{\em
  12,\/}~1-2, 89--116.

\bibitem[\protect\citeauthoryear{Gelfond and Lifschitz}{Gelfond and
  Lifschitz}{1988}]{gellif88b}
{\sc Gelfond, M.} {\sc and} {\sc Lifschitz, V.} 1988.
\newblock The stable model semantics for logic programming.
\newblock In {\em Proceedings of the Fifth International Conference and
  Symposium of Logic Programming (ICLP'88)}, {R.~Kowalski} {and} {K.~Bowen},
  Eds. MIT Press, 1070--1080.

\bibitem[\protect\citeauthoryear{Gelfond and Lifschitz}{Gelfond and
  Lifschitz}{1991}]{GL91}
{\sc Gelfond, M.} {\sc and} {\sc Lifschitz, V.} 1991.
\newblock Classical negation in logic programs and disjunctive databases.
\newblock {\em New Generation Computing\/}~{\em 9}, 365--385.

\bibitem[\protect\citeauthoryear{Gelfond and Przymusinska}{Gelfond and
  Przymusinska}{1993}]{GelfondP93}
{\sc Gelfond, M.} {\sc and} {\sc Przymusinska, H.} 1993.
\newblock Reasoning on open domains.
\newblock In {\em Logic Programming and Non-monotonic Reasoning, Proceedings of
  the Second International Workshop, Lisbon, Portugal, June 1993}, {L.~{Moniz
  Pereira}} {and} {A.~Nerode}, Eds. {MIT} Press, 397--413.

\bibitem[\protect\citeauthoryear{Kahl, Watson, Balai, Gelfond, and Zhang}{Kahl
  et~al\mbox{.}}{2015}]{kawabagezh15}
{\sc Kahl, P.}, {\sc Watson, R.}, {\sc Balai, E.}, {\sc Gelfond, M.}, {\sc and}
  {\sc Zhang, Y.} 2015.
\newblock The language of epistemic specifications (refined) including a
  prototype solver.
\newblock {\em Journal of Logic and Computation\/}.

\bibitem[\protect\citeauthoryear{Kelly}{Kelly}{2007}]{Kelly07}
{\sc Kelly, M.} 2007.
\newblock {Wviews}: A worldview solver for epistemic logic programs.
\newblock Honour's thesis (supervised by Yan Zhang). University of Western
  Sydney.

\bibitem[\protect\citeauthoryear{Leclerc and Kahl}{Leclerc and
  Kahl}{2018}]{leckah18}
{\sc Leclerc, A.} {\sc and} {\sc Kahl, P.} 2018.
\newblock A survey of advances in epistemic logic program solvers.
\newblock In {\em Proceedings of the Eleventh International Workshop on Answer
  Set Programming and other Computer Paradigms {(ASPOCP'18)}}.

\bibitem[\protect\citeauthoryear{Shen and Eiter}{Shen and
  Eiter}{2017}]{sheeit17a}
{\sc Shen, Y.} {\sc and} {\sc Eiter, T.} 2017.
\newblock Evaluating epistemic negation in answer set programming (extended
  abstract).
\newblock In {\em Proceedings of the Twenty-sixth International Joint
  Conference on Artificial Intelligence (IJCAI'17)}, {C.~Sierra}, Ed.
  IJCAI/AAAI Press, 5060--5064.

\bibitem[\protect\citeauthoryear{Son, Le, Kahl, and Leclerc}{Son
  et~al\mbox{.}}{2017}]{SLKL17}
{\sc Son, T.~C.}, {\sc Le, T.}, {\sc Kahl, P.~T.}, {\sc and} {\sc Leclerc,
  A.~P.} 2017.
\newblock On computing world views of epistemic logic programs.
\newblock In {\em Proc. of the 26th International Joint Conference on
  Artificial Intelligence, {IJCAI} 2017, Melbourne, Australia, August 19-25,
  2017}, {C.~Sierra}, Ed. ijcai.org, 1269--1275.

\bibitem[\protect\citeauthoryear{Truszczynski}{Truszczynski}{2011}]{truszczynski11b}
{\sc Truszczynski, M.} 2011.
\newblock Revisiting epistemic specifications.
\newblock {M.~Balduccini} {and} {T.~Son}, Eds. Lecture Notes in Computer
  Science, vol. 6565. Springer, 315--333.

\end{thebibliography}

\end{document}